\documentclass[aps,groupedaddress,floatfix,nofootinbib,preprintnumbers,eqsecnum,twocolumn]{revtex4}
\usepackage{amssymb,amsmath,graphicx,multirow,epsf}
\usepackage{dcolumn}   
\usepackage{bm}        

\begin{document}

\title{Completing $h$ \\}
\author{Keith R. Dienes\footnote{E-mail address:  {\tt dienes@email.arizona.edu}}}
\affiliation{
     Department of Physics, University of Arizona, Tucson, AZ  85721  USA}

\begin{abstract}
Nearly a decade ago, the science community was introduced to the $h$-index,
a proposed statistical measure of the collective impact of 
the publications of any individual researcher.
Of course,
any method of reducing a complex data set to a single number will
necessarily have certain limitations and introduce certain biases.
However, in this paper we point out that
the definition of the $h$-index
actually suffers from something far deeper:  a hidden mathematical incompleteness
intrinsic to its definition.
In particular, we point out that one critical step within the definition of $h$ 
has been missed until now, resulting in an index which only achieves
its stated objectives under certain rather limited circumstances.
For example, this incompleteness explains why the $h$-index ultimately has more utility
in certain scientific subfields than others.
In this paper, we expose the origin of this incompleteness and then also propose 
a method of completing the definition of $h$
in a way which remains close to its original guiding principle.
As a result, our ``completed'' $h$ not only reduces to the usual $h$ in cases where
the $h$-index already achieves its objectives, but also extends the validity of the 
$h$-index into situations where it currently does not.
\end{abstract}


\maketitle

\newcommand{\newc}{\newcommand}
\newc{\gsim}{\lower.7ex\hbox{$\;\stackrel{\textstyle>}{\sim}\;$}}
\newc{\lsim}{\lower.7ex\hbox{$\;\stackrel{\textstyle<}{\sim}\;$}}
\makeatletter
\newcommand{\biggg}{\bBigg@{3}}
\newcommand{\Biggg}{\bBigg@{4}}
\makeatother

\def\vac#1{{\bf \{{#1}\}}}

\def\beq{\begin{equation}}
\def\eeq{\end{equation}}
\def\beqn{\begin{eqnarray}}
\def\eeqn{\end{eqnarray}}
\def\calM{{\cal M}}
\def\calV{{\cal V}}
\def\calF{{\cal F}}
\def\half{{\textstyle{1\over 2}}}
\def\quarter{{\textstyle{1\over 4}}}
\def\ie{{\it i.e.}\/}
\def\eg{{\it e.g.}\/}
\def\etc{{\it etc}.\/}
\def\qbar{{ \overline{q} }}
\def\Nbar{{ \overline{N} }}
\def\chibar{{ \overline{\chi} }}
\def\psibar{{ \overline{\psi} }}


\def\inbar{\,\vrule height1.5ex width.4pt depth0pt}
\def\IR{\relax{\rm I\kern-.18em R}}
 \font\cmss=cmss10 \font\cmsss=cmss10 at 7pt
\def\IQ{\relax{\rm I\kern-.18em Q}}
\def\IZ{\relax\ifmmode\mathchoice
 {\hbox{\cmss Z\kern-.4em Z}}{\hbox{\cmss Z\kern-.4em Z}}
 {\lower.9pt\hbox{\cmsss Z\kern-.4em Z}}
 {\lower1.2pt\hbox{\cmsss Z\kern-.4em Z}}\else{\cmss Z\kern-.4em Z}\fi}
\def\OmegaDM{\Omega_{\mathrm{CDM}}}
\def\Omegatot{\Omega_{\mathrm{tot}}}
\def\rhocrit{\rho_{\mathrm{crit}}}
\def\arcsinh{\mbox{arcsinh}}
\def\BRgamma{\mathrm{BR}_{\lambda}^{(2\gamma)}}
\def\OmegaDM{\Omega_{\mathrm{CDM}}}
\def\tnow{t_{\mathrm{now}}}
\def\Omegatotnow{\Omega_{\mathrm{tot}}^\ast}
\def\erf{\mathrm{erf}}
\def\rhototloc{\rho^{\mathrm{loc}}_{\mathrm{tot}}}
\def\Ecut{E_{\mathrm{cut}}}
\def\Emax{E_{\mathrm{max}}}

\newcommand{\ipb}{\text{pb}^{-1}}
\newcommand{\ifb}{\text{fb}^{-1}}
\newcommand{\iab}{\text{ab}^{-1}}
\newcommand{\ev}{\text{eV}}
\newcommand{\kev}{\text{keV}}
\newcommand{\mev}{\text{MeV}}
\newcommand{\gev}{\text{GeV}}
\newcommand{\tev}{\text{TeV}}
\newcommand{\pb}{\text{pb}}
\newcommand{\mb}{\text{mb}}
\newcommand{\cm}{\text{cm}}
\newcommand{\m}{\text{m}}
\newcommand{\km}{\text{km}}
\newcommand{\kg}{\text{kg}}
\newcommand{\g}{\text{g}}
\newcommand{\s}{\text{s}}
\newcommand{\yr}{\text{yr}}
\newcommand{\Mpc}{\text{Mpc}}
\newcommand{\etal}{{\em et al.}}
\newcommand{\ibid}{{\em ibid.}}

\newcommand{\be}{\begin{equation}}
\newcommand{\ee}{\end{equation}}
\newcommand{\ba}{\begin{align}}
\newcommand{\ea}{\end{align}}

\hyphenation{ALPGEN}
\hyphenation{EVTGEN}
\hyphenation{PYTHIA}

\def\ie{{\it i.e.}\/}
\def\eg{{\it e.g.}\/}
\def\etc{{\it etc}.\/}
\def\Yeq{Y^{\mathrm{eq}}}
\def\peff{p_{\mathrm{eff}}}
\def\Weff{W_{\mathrm{eff}}}
\def\OmegaDM{\Omega_{\mathrm{DM}}}
\def\rhocrit{\rho_{\mathrm{crit}}}
\def\snow{s_{\mathrm{now}}}
\def\tnow{t_{\mathrm{now}}}
\def\wtM{\widetilde{M}}


\input epsf





\section{Introduction\label{sec:intro}}


In 2005, J.$\,$E.\ Hirsch introduced the so-called ``$h$-index'' as a way of assessing and quantifying
the impact of the publication record associated with an individual researcher~\cite{Hirsch,Hirsch2}.
Succinctly put, $h$ is defined as the number of papers that the individual in question
has produced
which have at least $h$ citations.
Phrased less succinctly but perhaps more usefully, 
$h$ is the maximum value of $N$ for which it can be said that the individual has 
$N$ papers with at least $N$ citations each.
The original motivation behind the definition of 
this index is that it balances between two opposite
poles:  the Scylla of total citation counts and the Charybdis of total numbers of papers.
Although the quotient of these two numbers (the average number of citations per paper)
is a useful measure for some purposes, it says nothing about how the citations are
actually distributed amongst the papers --- \ie, whether they are all associated with
just a few highly-cited papers, or whether they are distributed fairly evenly 
across the publications, with no single publication attracting
particularly strong attention. 
The $h$-index was therefore proposed as an alternative way of balancing between these
two extremes and thereby assessing the overall ``impact'' of a given publication record.

It goes without saying that 
any statistical method of reducing a complex data set to a single number will
necessarily have certain limitations that favor some researchers at the expense
of others.
Legitimate arguments can then be made for or against the proposed methodology,
and in the case of the $h$-index a large literature devoted to this topic already exists. 

It is not the purpose of this paper to engage in such discussions.
Rather, in this paper we wish to point out that
the definition of the $h$-index
actually suffers from something far deeper:  
a hidden mathematical incompleteness
intrinsic to its definition.
In particular, we will demonstrate that one critical step within the definition of $h$
has been missed until now, resulting in an index which only achieves
its stated objectives under the rather limited circumstances in which the missing
piece would not have had any effect.
However, in other cases, it turns out 
that this missing piece is responsible for the apparent failure 
of $h$ to act as originally desired.
For example, we shall see that 
this incompleteness explains why the $h$-index apparently has more utility
in certain scientific subfields than others.

Given this incompleteness in the definition of $h$,
we then take the next step and propose a method
of restoring the missing ingredient
in a manner which remains consistent with the original guiding principles underlying $h$.
As we shall see, this results in
a new, ``completed'' version of the $h$-index,
one which is mathematically robust across a wide variety of situations.
Of course, our ``completed'' $h$ reduces to the usual $h$ in cases where
the $h$-index already achieves its stated objectives.
However, more importantly, our ``completion'' of $h$ 
also extends its validity 
into situations where it currently does not.

\section{Exposing the problem with {\lowercase{\large\it h}}\/:\protect\\
  A simple scaling argument}

As described above, 
the $h$-index is designed to represent
a rather ingenious {\it balancing}\/ between paper counts and citation counts.
Rather than focus exclusively on either total numbers of papers or total numbers of citations,
$h$ looks at how the set of citations is actually distributed across the set of papers,
assessing the overall impact of a given publication record
by seeking the point at which 
the number of well-cited papers 
matches the minimum number of citations those papers have.
This balancing between paper counts and citation counts is the 
underlying motivation for $h$ as well as
the source of its ultimate utility.
Unlike other proposed assessment variables, $h$
is powerful because it represents neither variable exclusively
but instead relies upon a subtle comparison of the two against
each other.

However, it is easy to envision scenarios in which
this balancing fails --- \ie, situations in which  $h$ ends
up describing either a paper count {\it or}\/ a citation count,
with a value which is sensitive to only one of these variables
and essentially insensitive to the other.
For example, let us imagine two hypothetical
scientists:  one with $20$ papers whose citation
counts range from $1$ to $20$, and
one with $20$ papers whose citation
counts range from $100$ to $2000$.
In each case, the range of citation counts spans
a factor of $20$, and indeed 
the $h$-index of the first scientist
is smaller than that of the second, as expected.
However, we immediately see that
the $h$-index of the second scientist reduces to a mere paper count, 
in the sense that further citations will have absolutely no effect on his
$h$-index.
By contrast, this will generally not be the case for the first scientist.
  
Although this example is trivial,
it exposes the fact that the balancing inherent in $h$  ---
indeed, its uniquely valuable feature ---
is vulnerable to situations in which
paper counts and citation counts are of different orders
of magnitude.
In such cases, $h$ entirely loses its sensitivity
to one of these measures, and merely reflects the other.
In such cases, the $h$-index has failed in its primary purpose,
and no longer measures the subtle mixture
of variables 
it was designed to assess.

Of course, the situation described above is somewhat contrived and unrealistic. 
Perhaps the most unrealistic aspect of the above example 
is the fact that every paper of our second hypothetical scientist has 
a citation count which exceeds his
total number of papers.
This is extremely rare, if it ever happens at all ---
in general, the citation counts achieved by a given scientist will
range from some maximum value all the way down to zero.
Indeed, implicit in the original definition of the $h$-index is the assumption that
a given publication record 
will contain papers with numbers of citations both above and below $h$.

\begin{figure*}[t]
\begin{center}
  \epsfxsize 7.0 truein \epsfbox {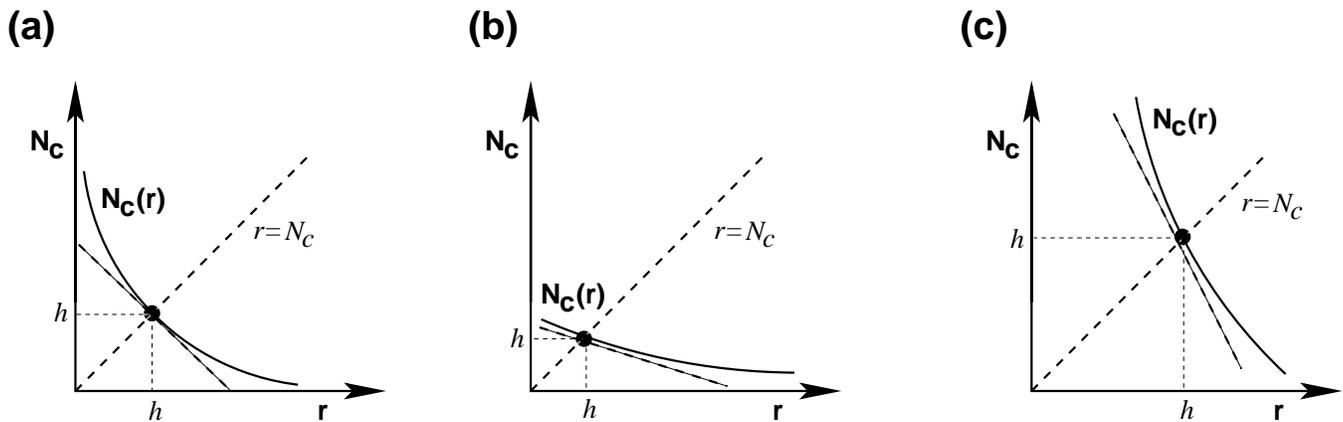}
\end{center}
\caption{Calculation of the traditional $h$-index, assuming a citation profile $N_c(r)$ 
as a function of paper rank $r$.
(a)  The scenario originally envisioned (and sketched) in Ref.~\cite{Hirsch}.
      It is assumed that the tangent to the $N_c(r)$ curve at $r=N_c(r)=h$ is
      perpendicular to the $r=N_c$ line, so that $h$ is equally balanced in
      its sensitivity to variations in numbers of papers versus numbers of citations.
(b,c)  Effects that occur when the numbers of citations are not commensurate with
       the numbers of papers, as can be achieved by rescaling the citation profile $N_c(r)$.
       In such cases, the tangent line for the $N_c(r)$ curve
       at $r=N_c(r)=h$ is no longer
       perpendicular to the $r=N_c$ line, implying that the value of the
       traditional $h$-index becomes extremely sensitive to variations of one variable while
       losing sensitivity to variations of the other.  }
\label{fig:spacetime}
\end{figure*}

However, even under these more restrictive conditions,
the overall scale associated
with citation counts can still have the effect of destroying the balance inherent
in $h$, thereby rendering $h$ essentially insensitive to one variable or the other.
To understand how this occurs,
    let us imagine ordering the publications of a given individual
    according to their citation rank $r$, so that the $r=1$ paper
    has the most citations and papers with increasing $r$-indices have 
    numbers of citations which either remain constant or decrease.
    Let us also assume that $N_c(r)$ represents the number of citations
    for each paper as a function of its rank $r$.
    In Fig.~\ref{fig:spacetime}, we have illustrated the graphical means
    by which the corresponding $h$-index may be calculated:  we simply 
    calculate the point at which the $N_c(r)$ curve intersects the $r=N_c$ line.
    Fig.~\ref{fig:spacetime}(a) illustrates the situation originally envisioned
    in Ref.~\cite{Hirsch}, where indeed an almost identical figure appears:  
    the overall scales for $r$ and $N_c$ are commensurate, so that the tangent
    line for the $N_c(r)$ curve is approximately perpendicular to the $r=N_c$ 
    line.  This implies that the resulting $h$-index represents  
    a true balancing between numbers of papers and numbers of citations. 
     In other words, the resulting $h$-index is just as sensitive 
   to variations in the citation counts $N_c$ as it is to variations
  in the paper rank $r$ (as would occur if further well-cited papers were produced).

    By contrast, in Figs.~\ref{fig:spacetime}(b) and 
    \ref{fig:spacetime}(c),
    we illustrate what occurs when the numbers of papers and  
    the numbers of citations are of different overall magnitudes.
    Indeed, all we have done in passing from
    Fig.~\ref{fig:spacetime}(a) to
    Figs.~\ref{fig:spacetime}(b) and \ref{fig:spacetime}(c)
    is to rescale the overall $N_c(r)$ curve by an arbitrary small or large numerical factor.
    As evident from 
    Figs.~\ref{fig:spacetime}(b) and \ref{fig:spacetime}(c),
    this has the effect of rescaling the corresponding {\it slope}\/ of
    the $N_c(r)$ curve at $r=h$ by the same factor.
    As a result, the tangent line for the $N_c(r)$ curve
       at $r=N_c(r)=h$ is no longer
       perpendicular to the $r=N_c$ line.  Indeed, for particularly
   small or large rescalings of the $N_c(r)$ curve [as illustrated
 in Fig.~\ref{fig:spacetime}(b) or 
 \ref{fig:spacetime}(c), respectively],
   the slope of the tangent line at $r=N_c(r)=h$ tends toward
   either zero or (negative) infinity.
   In such cases, $h$ becomes virtually insensitive to variations in
    either ranks or citation counts respectively. 

    This sensitivity issue is ultimately critical if $h$ is to retain
    its original intended meaning.  For example, 
    if the $N_c(r)$ curve has nearly vanishing slope at $r=N_c=h$,
    as in Fig.~\ref{fig:spacetime}(b),
    then the values of $N_c(r)$ with $r\gsim h$
    will not be too different from $N_c(h)$.
    Consequently it will only take a sprinkling of relatively few additional citations to raise
    the corresponding $h$-index significantly.  
    In other words, $h$ will be extremely sensitive to small variations in citation counts.
    By contrast, if the 
    $N_c(r)$ curve has a very steep slope at $r=N_c=h$,
    as in Fig.~\ref{fig:spacetime}(c),
    then the values of $N_c(r)$ with $r\gsim h$
    will be significantly smaller than $N_c(h)$.
    As a result, a moderate number of additional citations for these papers
    will not affect the value of the $h$-index:  $h$ 
     becomes relatively insensitive to small variations in citation counts.

To quantify these effects,
let us imagine increasing the citation counts by a small amount $\delta N_c$,
so that the entire $N_c(r)$ curve shifts according to 
\beq
      N_c(r) ~\to~ N'_c(r) \equiv N_c(r) + \delta N_c~.
\label{variation1}
\eeq
Alternatively, we could instead imagine 
a small variation in the publication counts,
as would occur if our hypothetical scientist were
to produce a small number $\delta r$ of additional highly-cited papers:
\beq
     N_c(r) ~\to~ N'_c(r) \equiv N_c(r-\delta r)~.
\label{variation2}
\eeq
Each of these actions will result in a shift $\delta h$ in the value of $h$ which solves
the defining equation $N_c(h)=h$.
In the first case, we find
\beq
       \delta h_1 ~\approx ~ {\delta N_c\over 1-x}
\label{find1}
\eeq
where
\beq  
          x~\equiv~ {d N_c(r)\over dr}{\bigg|}_{r=h}~<~0~ , 
\label{find2}
\eeq
while in the second case we find
\beq
       \delta h_2 ~\approx ~ \left( {{-x }\over {1-x}}\right)\, \delta r~.
\label{find3}
\eeq
As a result, setting $\delta h_1 = \delta h_2$, we see that 
these variations will have equal (``balanced'') effects on $h$ 
only if
\beq
      \delta N_c ~=~ - x\, \delta r~.
\label{find4}
\eeq

If $|x|\gg 1$ [as in Fig.~\ref{fig:spacetime}(c)],
we thus see that $\delta N_c$ must be significantly larger than $\delta r$ in order
to have the same effect on $h$ --- \ie, $h$ has become essentially independent of the citation counts $N_c$.
By contrast, if $|x|\ll 1$ [as in Fig.~\ref{fig:spacetime}(b)],
the opposite is true, and $\delta N_c$ becomes extremely small compared with $\delta r$ ---
\ie, $h$ has become overwhelmingly sensitive to the citation counts.
It is only for $|x|\approx 1$ that a true balance is achieved, with 
$h$ exhibiting a roughly equal sensitivity to variations in paper counts and citation counts,
and indeed it is this situation
which was implicitly assumed in Ref.~\cite{Hirsch}.

Our discussion thus far has illustrated 
the problems that arise when citation counts $N_c$ are rescaled.
But why should we care about rescalings of $N_c$?

To understand this issue  ---
and to see why this problem is particularly dangerous for the usefulness of $h$
as a measure of impact ---
let us do a thought experiment and 
restrict our attention to a certain community of scientists.
In general we shall consider a ``community'' to be any collection
of scientists who tend to draw citations from and bestow citations upon
each other;    
for example, we might consider the community of theoretical high-energy physicists,
or the community of experimental condensed-matter physicists.
We shall refer to this as Community~$A$.
    Let us further assume that within this 
    community, the average good scientist writes $N^{(A)}_p$ papers, and 
    that each good paper receives an average of $N^{(A)}_c$ citations.
    As long as $N_p^{(A)}$ and $N_c^{(A)}$ are of the same rough order of magnitude,
    the $h$-index will measure what it was designed to measure,
    balancing two relevant measures of the 
    output of the majority of the scientists in this community.
 
    But now let us imagine 
    a second community $B$ which can be considered to be an ensemble
    of ten identical $A$ communities.  Schematically, we shall write $B=10 A$.
    Each scientist in Community~$B$
    is exactly as productive as he was before, when he was simply a member of
    Community~$A$, and thus
    the number of papers that he might write in his career is unaffected
    by his transition from Community~$A$ to
    Community~$B$.~  Thus his number of papers is invariant under the scaling of
    the community from $A$ to $B$:  \ie, we have $N_p^{(B)}=N_p^{(A)}$.
    But if the community grows from $A$ to $B$, and
    if each community within the ensemble is identical, one would now 
    expect this scientist to receive 10 times the number of citations
    for each of his papers as he would have received from 
    Community~$A$ alone:  $N_c^{(B)}= 10 N_c^{(A)}$.
    In other words, we see that $N_c$ --- unlike $N_p$ ---
    scales with community size.
    Thus, if a particular balancing $N_p^{(A)} \sim N_c^{(A)}$ happened to hold within Community~$A$, 
    then $N_p^{(B)} \not \sim N_c^{(B)}$ --- \ie, this balancing
    will no longer hold for the members of Community~$B$.
    Indeed, the latter scientists will find that
    their numbers of citations
    will scale out of proportion to their number of papers.  Or, phrased
    more precisely, the $N_c(r)$ curves of scientists in Community $B$ will
    be scaled relative to the $N_c(r)$ curves of scientists in Community $A$.
    Thus, for scientists in Community~$B$,
    the $h$-index will be reduced to a mere measure of their paper counts
    and show almost no sensitivity to citation counts.
    Indeed, this will happen for each member of Community~$B$.
    Thus, within Community~$B$,
    the $h$-index fails to assess what it was designed to assess. 

    As discussed above, this conclusion rests upon our assertion that $N_c$ scales
    with community size.
      This is a direct consequence of our original supposition that Community~$B$ is an
      ensemble of identical copies of Community~$A$.~  Note that this also directly implies 
      that each paper published within Community~$B$ will have a reference list which is
      10 times longer than those within Community~$A$ --- a fact which, though true, is
      not relevant for our purposes.  In reality, of course, one might expect 
      that the papers published by scientists in Community~$B$ will each
      experience an increased 
      competition to be noticed --- an effect which might tend to suppress the 
      citation counts within Community~$B$ relative to the pure-ensemble result.
      However, given that there is 
      no upper limit on the numbers of references one can have in a given paper,
      it is our opinion (though apparently not that of the author of Ref.~\cite{Hirsch}) 
      that this suppression effect is certainly subleading.
     Thus, even if $N_c^{(B)}$ is not exactly equal to $10 N_c^{(A)}$, 
      we nevertheless expect to find that $N_c^{(B)}\gg N_c^{(A)}$.  
      Our main conclusion thus still holds:
      the $h$-index will tend to fail within Community~$B$ even
    if it functioned perfectly well within Community~$A$.

    This is problematic because a given scientist has no control over the
    size of his or her community.  The size of the community is a feature
    which is {\it external}\/ to the scientist in question, yet we see
    that this feature has the disastrous effect of determining
    whether his or her $h$-index is functioning as designed, as a balanced
    measure of scientific impact.
    This is particularly distressing because our supposition that Community~$B$
    is an ensemble of identical copies of 
    Community~$A$ guarantees that the scientist in Community~$A$
    is identical to the scientist in Community~$B$.

    It is, of course, to be expected 
    that certain measures associated
    with an individual publication record 
    (such as citation count)
    will scale with community size,
    while others (such as paper count) will not.   
    Indeed, neither type of scaling behavior causes particular difficulties when 
    one is comparing individuals within a given community because
    each of these measures (\eg, paper count or citation count) 
    is subjected to a uniform scaling behavior.  The same is
    even true of their ratio.  However, the $h$-index is intrinsically different
    from these other kinds of measures because it aims at something different:  
    it is meant to be a {\it comparison}\/ between two separate measures, 
    one of which scales with community size and the other not.
    Thus, while $h$ may have relevance for one community of scientists
    in which we might expect $N_p\sim N_c$, 
    it may completely fail to have any relevance for another in which
    $N_p$ and $N_c$ are expected to be grossly dissimilar.

    We can state this problem more mathematically as follows.
    Both $N_p$ (publication counts) and $N_c$ (citation counts) transform
    {\it covariantly}\/ with respect to such rescalings --- indeed, $N_p$ is actually invariant,
    transforming as $s^0$ where $s$ is the rescaling factor, while $N_c$ is only covariant,
    transforming as $s^1$.
    Even their quotient (the average number of citations per paper) 
    transforms covariantly, as $s^1$.
    Unfortunately, the $h$-index does {\it not}\/ transform covariantly under
    rescalings of community size.

    Even worse, the changes in $h$ under such rescalings are
    actually {\it different}\/ for different scientists within the same community!
    To see this, let us imagine Scientist~$X$ who has 10 papers, each with 5 citations,
    and Scientist~$Y$ who has 10 papers, each with 10 citations.
    Clearly $h_X=5$ and $h_Y=10$.
    But let us now rescale the community size by a factor of 2.
    Scientist~$X$ will now have $10$ papers, each with $10$ citations,
    while Scientist~$Y$ will now have $10$ papers, each with $20$ citations.
    Thus, after the rescaling, we now find that $h_X=h_Y=10$.
    Amazingly, two scientists who originally had very different $h$-indices 
    will now be deemed equally meritorious --- all while nothing changed except
    the number of {\it other}\/ scientists in the community!
    Indeed, this is only one of a number of such practical inconsistencies inherent
    in the traditional $h$-index;  other similar inconsistencies are
    discussed in Refs.~\cite{vanEck,waltman}.
      
    Ultimately, then, we see that the strength of $h$ is also its weakness:
    it compares numbers of papers with numbers of citations,
    yet these are different things.
    This general point has also been emphasized in, \eg, Refs.~\cite{lehmann,lehmann2,vanEck,waltman}.
    More specifically, these two quantities (paper counts and citation counts)
    transform differently under
    community rescalings.  Thus, each can in general carry a different
    intrinsic scale associated with it.
    When these two scales are similar, the $h$-index has meaning.
    However, when these scales are dissimilar, the $h$-index loses
    that feature which makes it unique, and instead develops
    an overwhelming and ultimately misleading sensitivity to small variations 
    in either paper counts or citation counts.
    The $h$-index then no longer functions as originally envisioned.

\section{Dimensional analysis to the rescue:~
        Completing the definition for {\lowercase{\large\it h}}}

Thus far, we have discussed some of the symptoms that indicate 
that something is amiss with $h$.
However, we have not yet discussed the underlying disease.
    
Phrased in simple terms, 
these problems with $h$ are traceable to a simple underlying source:
the definition of $h$ ultimately suffers from a ``dimensional'' inconsistency.
If the definition of the $h$-index had concentrated purely on numbers of citations
or numbers of papers, all would have been well.  Even their quotient would
not have been problematic.
However, the $h$-index reaches further and attempts something unique:  a
direct comparison between a certain number of papers and a certain number of
citations.
It is, of course, certainly true that a number of papers and a number of citations both
have the same dimensionality:  they are both pure numbers.
But this is too simplistic.

We readily accept, for example, that length and time
are quantities with different dimensionalities:  for example,  one
is measured in meters and the other in seconds.
As a result, we do not compare the magnitude of a length interval with the
magnitude of a time interval.
Indeed, in this context it is important to note that
we even refrain from engaging in the verbal
gymnastics of trying to compare the 
 {\it number}\/ of meters associated with a given length interval
with the {\it number}\/
of seconds associated with a given time interval, even though both are now pure numbers.
Rather, we can only compare these two quantities when
we have a reference conversion factor --- in this case,
a relevant fiducial velocity, such as a speed of light $c$.
Indeed, it is only by having the speed of light to 
serve as a conversion factor that such a comparison
can be made.

The same is true for our comparison between papers and citations.
Papers and citations are different things, essentially behaving as
quantities with different dimensionalities.
Unfortunately, the standard approach to $h$ compares them directly.
In so doing, this procedure implicitly sets $c=1$ within
our Euclidean $(r,N_c)$ ``spacetime''.
This is reflected in the ``spacetime'' diagrams in Fig.~\ref{fig:spacetime}, wherein
the $h$-defining condition $r=N_c(r)$ is always a line with fixed angle of 45$^\circ$ relative
to the horizontal.

Setting $c=1$ is certainly useful for many purposes, and high-energy physicists
do this quite frequently.  
However, we cannot keep the speed of light fixed at $c=1$ if we wish to consider
rescalings of lengths relative to times, since this missing velocity factor carries a lot
of hidden theoretical scaling information that would otherwise be lost.
Even more urgently, we cannot set $c=1$ if we wish to calculate actual physical
 {\it numbers}\/;  we must instead use the correct numerical value for $c$, 
written in terms of meters and seconds. 
Unfortunately, the standard definition of $h$ misses this point entirely,
and thereby implicitly takes this conversion factor as unity as a general statement.
This is an error of dimensional analysis, and directly leads to all of the scaling 
difficulties observed above.

Our recipe for completing $h$ is therefore obvious:  we must restore a 
missing dimensionful conversion factor ---
a missing ``speed of light'' --- into the comparison between paper counts
and citation counts.
Specifically, for any scientist within a given community~$A$ who has 
a citation profile described by a function $N_c(r)$, 
we define $h$ as the solution to the condition
\beq
         N_c(h) ~=~ c_A \, h~.
\label{cond1}
\eeq
Clearly, as a ``velocity'' within our $(r,N_c)$ ``spacetime'',
the quantity $c_A$ will have dimensions of citations per paper.
As we shall see, introducing this missing factor will also solve the 
problematic scaling issue discussed above, and likewise render the resulting
$h$-index covariant with respect to rescalings in community size.
In this connection, we note that the possibility of adopting an equation
such as Eq.~(\ref{cond1}) 
was also discussed in Refs.~\cite{vanEck,waltman}
in order to illustrate some of the fine-tuned arbitrariness 
inherent in the traditional definition of $h$.
Similar work along these lines also appears 
in Refs.~\cite{lehmann,lehmann2,Komulski,Wu,Ellison,Schreibermore,Schreiberplus}.
However, our main point here is that adopting such a definition
is not merely a {\it possibility}\/ but rather a
logical and mathematical {\it necessity}\/.

The only question, then, is to determine the correct numerical value of $c_A$.
In order to do this, we 
make recourse to the original guiding principle that underlies the $h$-index:
we choose $c_A$ such that the resulting $h$-index 
compares citation counts against paper counts in a way that
balances the two against each other, with equal sensitivities to
variations in these two quantities.
Of course, there is no magic value of $c_A$ which ensures that every member of
Community~$A$ will have an $h$-index which is properly balanced in this way.
Indeed, each individual will have a presumably unique citation function
$N_c(r)$, and we have already seen that proper balancing is sensitive to the
derivatives of this function.
On the other hand, a {\it community}\/ of scientists will have 
a ``collective'' citation function $N^{(A)}_c(r)$, where $N_c^{(A)}(1)$ is defined to be the
average citation count associated with the top-cited paper from each member of the community,
where $N_c^{(A)}(2)$ is the average citation count associated with the second-most cited paper
from each member of the community, and so forth.
While the individual citation profile $N_c(r)$ from each member of 
the community may vary significantly and exhibit a somewhat jagged, irregular behavior,
we expect that the {\it collective}\/ citation count $N_c^{(A)}(r)$ across a fairly large community
will exhibit a relatively smooth behavior.  As a result, it should be fairly straightforward
(at least numerically) to evaluate the derivatives of this function.

Our procedure is then clear:  we simply evaluate an appropriate ``speed of light'' $c_A$
to be associated with Community~$A$ in such a way that on average, the corresponding $h$-indices
of the members of this community come as close as possible to having
equal sensitivities to variations in citation counts and paper counts.
Operationally, if we imagine a single individual who collectively represents the
community in the sense that his personal citation profile
exactly matches the collective profile $N_c^{(A)}(r)$ of the community, we wish to calculate 
an appropriate ``speed of light'' $c_A$ for this
individual so that his $h$-index $h_A$ will be properly balanced.
This will then define $c_A$ for the community he represents.

Our analysis proceeds as before.
Given the collective community citation profile $N_c^{(A)}(r)$,
we of course require 
\beq
   N_c^{(A)}(h_A) ~=~ c_A \, h_A~.
\label{con1}
\eeq
However, we also require that our solution for $h_A$ be properly balanced.
Shifts of the form $N_c^{(A)}(r) ~\to~ N_c^{(A)}(r) + \delta N_c$
will result in variations of the form
\beq
            \delta h_A ~\approx~ {\delta N_c\over c_A-x}
\eeq
where 
\beq 
          x~\equiv~ {dN_c^{(A)}(r) \over dr} \bigg|_{r=h_A}~,
\eeq
in complete analogy with Eqs.~(\ref{find1}) and (\ref{find2}).
Likewise, shifts of the form 
$N_c^{(A)}(r) ~\to~ N_c^{(A)}(r-\delta r)$
will result in variations of the form
\beq
           \delta h_A ~\approx~ \left({-x\over c_A -x}\right) \, \delta r~,
\eeq
in analogy with Eq.~(\ref{find3}).
Balanced effects therefore arise only when
the condition in Eq.~(\ref{find4}) continues to hold.
However, 
we want these effects to be naturally and automatically balanced when
the shifts $\delta r$ and $\delta N_c$ are
of the {\it same}\/ order of magnitude, as related through 
our ``speed of light'' conversion factor:  
\beq
          \delta N_c ~=~  \delta(c_A r)~.
\label{find5}
\eeq
Comparing Eqs.~(\ref{find4}) and (\ref{find5}), we therefore see that
there is only one way in which this can occur:
we require
\beq
        c_A ~= ~   -x ~\equiv ~ - {dN_c^{(A)}(r)\over dr}\bigg|_{r=h_A}~.
\label{con2}
\eeq
In other words, $c_A$ must be equal to (the negative of) the slope of the $N_c^{(A)}(r)$ curve
at $r=h_A$.
 
We thus are left with two equations, Eqs.~(\ref{con1}) and (\ref{con2}), which 
must be solved simultaneously for our two variables $c_A$ and $h_A$
using the community profile function $N_c^{(A)}(r)$.
The solution for $c_A$ then defines the 
``speed of light'' for Community~$A$,
whereupon we can easily determine the $h$-index for any individual
member of this community through the single defining equation in Eq.~(\ref{cond1})
using his/her own personal citation profile $N_c(r)$.

\begin{figure}[t]
\begin{center}
  \epsfxsize 2.5 truein \epsfbox {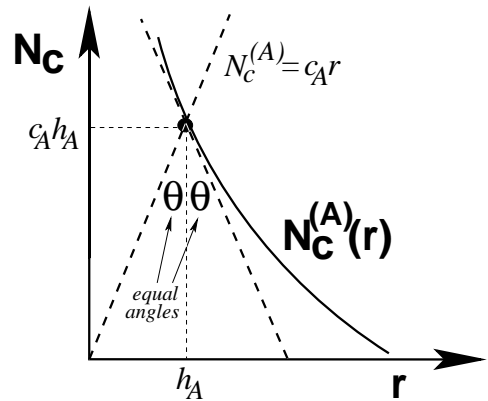}
\end{center}
\caption{Calculation of the ``speed of light'' factor $c_A$ for Community~$A$, given
    its average citation profile $N_c^{(A)}(r)$ as a function of paper rank $r$.
     We simply scan along the curve $N_c^{(A)}(r)$,
     seeking a location at which the angle $\theta$ made by the tangent with respect to the
    vertical matches the angle $\theta$ made by a line connecting that location to the   
    origin.  The ``speed of light'' is then given by $c_A = \cot\theta$.} 
\label{fig:spacetime2}
\end{figure}

The two constraint equations (\ref{con1}) and (\ref{con2}) are easy to interpret
graphically, as illustrated in Fig.~\ref{fig:spacetime2}.
Given the citation profile $N_c^{(A)}(r)$ as a function of paper rank $r$,
we simply scan along the curve $N_c^{(A)}(r)$,
seeking a location at which the slope of the tangent line is exactly equal
and opposite to the slope of a line connecting that location to the origin.
Of course, equal and opposite slopes imply that the two angles 
labelled $\theta$ in Fig.~\ref{fig:spacetime2}
will be equal.
Once such a location is identified, 
the ``speed of light'' for the community is then given by $c_A = \cot\theta$. 
Note that if $\theta$ turns out to be $\pi/4$, we have $c_A=1$;
this is then the original situation envisioned in Ref.~\cite{Hirsch} and
sketched in Fig.~\ref{fig:spacetime}(a).
However, in general, we find that $c_A\not=1$.
This means that for scientists in Community~$A$, 
the $N_c\approx r$ portions of their $N_c(r)$ curves will no longer be relevant for determining
their $h$-indices;  rather, it will be the $N_c\approx c_A r$ portions which now become relevant.

Our procedure for calculating $h$-indices 
can therefore be compared with that in Ref.~\cite{Hirsch} as follows.
In Ref.~\cite{Hirsch}, the value of $h$ is determined by solving a single constraint equation
\beq
         N_c(h) ~=~ h.
\eeq
Because one assumes a ``speed of light'' $c=1$ as a universal conversion factor, this can be done directly at the level
of the individual researcher, using his or her own citation profile $N_c(r)$. 
By contrast, our ``completed'' $h$-index is calculated in conjunction with a conversion factor $c$ by solving two  
constraint equations simultaneously:
\beq
        \begin{cases}
        &  N_c(h) ~=~ c\, h \cr
        &  N'_c(h)~=~ -c~, \cr
         \end{cases}
\label{simult}
\eeq
where $N'_c(r)\equiv dN_c(r)/dr$.
This is done at the level of the relevant community $A$, using the collective community
citation profile $N^{(A)}_c(r)$.  
This then defines the value $c_A$
for the community, whereupon each individual within the community is assigned an $h$-value 
according to the constraint $N_c(h) = c_A h$ using his/her own citation profile $N_c(r)$.

\begin{table*}
\begin{tabular}{||c||l|| c|c|c|c|c||}
\hline
\hline
 ~ & ~~~~~~~~~~citation profile & ~~$N_c(1)$~~ & ~~$r_{\rm max}$~~ 
                & ~~$N_c^{\rm (avg)}$~~& ~~traditional~$h$ ~~ &  ~~~$(h_A,  c_A)$~~~ \\
\hline
\hline
I &  ~~~$N_c(r)= 200 -3 r$ & $197$ & $66$ &  $99.5$  &  $h=50$ & 
          $\begin{matrix}
            h=33\cr  
            c=3\cr 
          \end{matrix}$ \\ 
\hline
II &  ~~~$N_c(r)= 200 -20 \sqrt{r} $ & $180$ & $99$ & $66.4$  &  $h=53$ & 
          $\begin{matrix}
            h=44\cr  
            c=1.5\cr 
          \end{matrix}$ \\ 
\hline
III &  ~~~$N_c(r)= 300 -180 \, r^{0.1} $ & $120$ & $159$ & $27.9$ & $h=39$ &   
          $\begin{matrix}
            h=63\cr  
            c=0.43\cr 
          \end{matrix}$ \\ 
\hline
IV &  ~~~$N_c(r)= 480 -300 \,r^{0.1} $ & $180$ & $107$ & $43.8$ & $h=43$ & 
          $\begin{matrix}
            h=42\cr  
            c=1.03\cr 
          \end{matrix}$ \\ 
\hline
V &  ~~~$N_c(r)= 200 \exp (-r/20)$ & $190$ & $105$ & $37.0$ & $h=34$ & 
          $\begin{matrix}
            h=20\cr  
            c=3.67\cr 
          \end{matrix}$ \\ 
\hline
VI &  ~~~$N_c(r)= 200 \exp [-(r/5)^{2/3}] $~~~ & $142$ & $60$ &  $20.3$   & $h=18$ & 
          $\begin{matrix}
            h=9\cr  
            c=4.86\cr 
          \end{matrix}$ \\ 
\hline
VII &  ~~~$N_c(r)= 300 \exp [-(r/3)^{3/5}] $~~~ & $178$ & $54$ & $22.4$   & $h=17$ & 
          $\begin{matrix}
            h=7\cr  
            c=8.06\cr 
          \end{matrix}$ \\ 
\hline
\hline
\end{tabular}
\caption{ ~Examples of different citation profiles $N^{(A)}_c(r)$.  For each we have calculated 
     the corresponding value of the traditional $h$-index that would have been ascribed to this profile as well as 
    our modified $h_A$-index and ``speed of light'' $c_A$ that result from solving the simultaneous
   equations (\ref{con1}) and (\ref{con2}).  
    It is clear that in general
   $c_A\not=1$.
   We have also listed the average number of citations per paper $N_c^{\rm (avg)}$ in each case, 
      calculating this quantity for the first $r_{\rm max}$ papers.
     }
\label{table1}
\end{table*}

Given an arbitrary citation profile function $N_c^{(A)}$,
it is not always guaranteed that a solution for $c_A$ exists.
However, for functions $N_c^{(A)}$ which are monotonically decreasing with non-positive second
derivatives (as are the most realistic for describing citation profiles),
we find that solutions for $h_A$ and $c_A$ can generally be found.
Examples are shown in Table~\ref{table1}, where we consider a linear function,
various power-law functions,
and several exponential functions
(both stretched and unstretched).
Indeed, it is shown in Refs.~\cite{stretchedexp,Hirsch}
that stretched exponential functions
of this form do a particularly good job
of modeling realistic citation profiles.
For each citation profile in Table~\ref{table1}, we have listed
the number of citations $N_c(1)$ associated with the most-cited paper,
the number $r_{\rm max}$ of papers which have
citation counts $N_c \geq 1$,
the average number of citations per paper 
$N_c^{\rm (avg)}$ when only the first $r_{\rm max}$ papers are included,
the traditional $h$-index that would have
been associated with this citation profile (\ie, the
traditional $h$-index of a scientist with this citation profile),
and the values of $h_A$ and $c_A$ corresponding
to this profile.

In all of these cases, we see from Table~\ref{table1} that solutions for $h_A$ and $c_A$
can indeed be found;  moreover, we find that $h_A$ generally differs significantly from the traditional
$h$-index that would have been obtained in each case.  Likewise, we see that $c_A$ generally differs from 
unity, often quite significantly.
Moreover, as a result of the shapes of these profiles,
we generally find that $c_A >1$ when $h_A < h$,
     and $c_A <1$ when $h_A >  h$.  
Note that while our main interest here is in the value of $c_A$, 
the value of $h_A$ also carries information that will be useful in what follows.

There are, however, certain situations in which no solutions for $c_A$ exist.  Likewise,
there are situations in which multiple solutions for $c_A$ can exist.
An extreme case demonstrating both features is the citation profile $N^{(A)}_c(r)=N_0/r^\beta$.
For $\beta=1$, we find that our two constraint equations in Eq.~(\ref{simult}) permit {\it any}\/  
value for $c_A$, including $c_A=1$.
Indeed, this is a very special citation profile, known as the Zipf distribution~\cite{Zipf}, which has 
been hypothesized to apply to a large variety of statistical systems dominated by essentially random behavior. 
We thus see that the Zipf distribution is also unique in that it happens to maintain a perfectly balanced
$h$-index regardless of the corresponding ``speed of light''!
By contrast, for $\beta\not=1$, we see that no solution for $c_A$ exists.
In such cases, our goal should be to find values of $c_A$ which come as close as possible to satisfying
the constraint equations within the allowed range $1\leq r \leq r_{\rm max}$.

As an illustration of our procedure for calculating the $h$-index, let us imagine Scientist~$X$
whose personal citation profile is given by $N^{(X)}_c(r)= 200 -3 r$,
and Scientist~$Y$ whose personal citation profile is given by 
$N^{(Y)}_c(r)= 300 \exp[-(r/3)^{3/5}]$.
Let us further imagine that both of these scientists are members of  
Community~$A$, whose collective community citation profile is given by
$N^{(A)}_c(r)= 200 \exp (-r/20)$.
Consulting Table~\ref{table1}, we see that the ``speed of light'' conversion factor 
for Community~$A$ is $c_A=3.67$.
We then set $N^{(X)}_c(h) =3.67 h$ to find that Scientist~$X$
has $h^{(X)}=29$, and set
$N^{(Y)}_c(h)= 3.67 h$ to find that Scientist~$Y$
has $h^{(Y)}= 10$.
By contrast, if we had arbitrarily set $c_A=1$ for these scientists,
we would have found $h^{(X)}=50$ and $h^{(Y)} = 17$.

Clearly, this procedure for calculating the $h$-index is more complex than the traditional one, 
and requires the extra knowledge of the
community in which the scientist is embedded.
Fortunately, at a practical level, 
such community citation profiles need to be calculated only once per community;
the ``speed of light'' is then universal for all scientists in that community.
Moreover, as discussed in Sect.~II, there are three critically important benefits to our method
which are lacking in the traditional procedure:
\begin{itemize}
  \item  On average, the $h$-indices assigned to scientists within the community
             are as close as possible to being ``balanced''.  Thus they function
              as anticipated, and are as close as possible to being equally sensitive 
             to variations in paper counts and citation counts.
             As we have discussed, this is generally 
            not true for the $h$-indices calculated by arbitrarily taking $c_A=1$.     
  \item  Likewise, our $h$-indices are {\it invariant}\/ under rescalings of the size
            of the community.  For example, in the above situation, let us imagine that 
            Scientists~$X$ and $Y$ are no longer members of Community~$A$ but
            rather members of the ``ensemble'' Community~$B=10A$, as discussed in Sect.~II.~
            Repeating our calculation but with $N_c^{(B)}(r)=10 N_c^{(A)}(r)$, 
            we continue to find $h^{(X)}=29$ and $h^{(Y)}= 10$, as desired;  indeed
            only the ``speed of light'' has changed, with $c_B=10 c_A$.
            By contrast, if we were to implement the traditional calculation with the ``speed of light'' 
            held fixed at $c=1$, 
            we would instead find
            $h^{(X)}= 64$ [up from its previous value $h^{(X)}= 50$]
               and 
            $h^{(Y)}= 35$ [up from its previous value $h^{(Y)}= 17$].
            As discussed in Sect.~II, this is an extremely undesirable outcome, 
            since these new $h$-indices are likely
            to be seriously unbalanced compared to the original ones, exhibiting great sensitivity to 
            small variations in paper counts at the expense of citation counts.
            Even worse, we see that this ten-fold magnification of the size of the community 
            would have an unequal effect on Scientists~$X$ and $Y$:  the former's $h$-index
            would rise by 28\% while the latter's would rise by a full 100\%!  
            It is only through the use of our modified $h$-index with a variable ``speed of light'' parameter 
            that we have been able to sidestep all of these problematic issues.
  \item  Finally, because our procedure assigns a different ``speed of light''
            conversion factor $c_A$ to different communities (thereby
           ensuring proper balancing within each community 
           as well as ensuring invariance under rescalings of community size),
            these modified $h$-indices can be more meaningfully compared 
            {\it across}\/ different communities than the traditional $h$-indices. 
            We shall discuss precisely how this may be done in more detail below, but the end
              result will be that we simply normalize the $h$-indices of the scientists in each community
              with respect to the $h_A$-index associated with that 
            community before performing an inter-community comparison.
            This ability to compare our ``completed'' $h$-indices across disciplines is 
            a particularly valuable feature, given that
            the magnitudes of the traditional $h$-indices are known to vary significantly from
            discipline to discipline.
\end{itemize}

\begin{table*}
\begin{tabular}{||c||l|| c|c|c|c|c||}
\hline
\hline
 ~ & ~~~~~~~~~~citation profile & ~~$N_c(1)$~~ & ~~$r_{\rm max}$~~ 
                & ~~$N_c^{\rm (avg)}$~~& ~~traditional~$h$ ~~ &  ~~~$(h_A,  c_A)$~~~ \\
\hline
\hline
I$'$ &  ~~~$N_c(r)= 202 - 2 r$ & $200$ & $100$ &  $102$  &  $h=67$ & 
          $\begin{matrix}
            h=50\cr  
            c=2\cr 
          \end{matrix}$ \\ 
\hline
II$'$ &  ~~~$N_c(r)= 222 -22 \sqrt{r} $ & $200$ & $100$ & $74.3$  &  $h=56$ & 
          $\begin{matrix}
            h=45\cr  
            c=1.63\cr 
          \end{matrix}$ \\ 
\hline
~III$'$, IV$'$~ &  ~~~$N_c(r)= 539.34 -339.34 \,r^{0.1} $ & $200$ & $100$ & $49.1$ & $h=43$ & 
          $\begin{matrix}
            h=39\cr  
            c=1.24\cr 
          \end{matrix}$ \\ 
\hline
V$'$ &  ~~~$N_c(r)= 210.9 \exp (-r/18.838)$ & $200$ & $100$ & $38.5$ & $h=34$ & 
          $\begin{matrix}
            h=18\cr  
            c=4.12\cr 
          \end{matrix}$ \\ 
\hline
VI$'$ &  ~~~$N_c(r)= 258.5 \exp [-(r/7.694)^{2/3}] $~~~ & $200$ & $100$ &  $25.0$   & $h=26$ & 
          $\begin{matrix}
            h=14\cr  
            c=4.08\cr 
          \end{matrix}$ \\ 
\hline
VII$'$ &  ~~~$N_c(r)= 285.3 \exp [-(r/5.613)^{3/5}] $~~~ & $200$ & $100$ & $22.5$   & $h=26$ & 
          $\begin{matrix}
            h=13\cr  
            c=4.10\cr 
          \end{matrix}$ \\ 
\hline
\hline
\end{tabular}
\caption{ ~The same citation profiles $N^{(A)}_c(r)$ as in Table~\ref{table1}
   except that their numerical coefficients have now been fine-tuned  
   in each case so as to produce $N_c(1)=200$ and $r_{\rm max}=100$.   In some sense, these 
   represent fixed top and bottom ``boundary conditions'' for our profiles;  
   thus, the only remaining differences between these profiles are 
   their different functional dependences on $r$ 
   between $r=1$ and $r=r_{\rm max}$.
   Data is shown for each profile, as in Table~\ref{table1}, and we observe the same general
   trends as in Table~\ref{table1}.
   The variations in the data across the different profiles
   in this table are thus purely the result of differences in the functional forms of these
   profiles between $r=1$ and $r=r_{\rm max}$,
   and are wholly independent of
   possible differences in their ``boundary'' values $N_c(1)$ and $r_{\rm max}$.}
\label{table2}
\end{table*}

In Table~\ref{table1}, we listed the values of 
the traditional $h$-index, the modified $h$-index,
and the ``speed of light'' $c_A$ for a variety of different citation profiles $N_c(r)$.
However, for some purposes 
it may also be of interest to study how these results vary across different profiles.
In general, there are two distinct ways in which two profiles $N_c(r)$ might differ.
First, they may differ
in their ``boundary'' values at $r=1$ and $r=r_{\rm max}$, where $r_{\rm max}$ is defined as above.
However, even if those boundary values are held fixed, two profiles 
may also differ in their overall functional forms.  For example, 
Profiles~III and IV in Table~\ref{table1} differ in the first way but not the second, while the rest differ
in both ways simultaneously.
In order to isolate the effects of these two kinds of differences, we can adjust the numerical coefficients
in each of the profiles listed in Table~\ref{table1} so as to bring all of these profiles to share a common
value of $N_c(1)$ and $r_{\rm max}$.
The resulting data is collected in Table~\ref{table2}.
As might be expected, we see that the variations in the data across the different profiles 
in Table~\ref{table2} are 
somewhat less sharp than they were in Table~\ref{table1}, when both the profile ``boundary'' values {\it and}\/
the profile functional forms were allowed to vary simultaneously. 
Nevertheless in each case we see that the traditional $h$-index continues to differ significantly from
the modified $h$-index, indicating that our proposed completion of the $h$-index continues to produce
a significantly different numerical outcome.

\section{Discussion}

Amongst the major statistical indices which seek to encapsulate
the impact of a given publication record,
the $h$-index is perhaps unique in that it seeks to balance
paper counts and citation counts against each other through
a direct numerical comparison.
This novel idea has led not only to an explosion of interest in the
properties of this index, but also to its rapid, near-universal adoption
within the scientific community.

Unfortunately, the fundamental mathematical aspect of this index ---
its direct numerical comparison between paper counts and citation counts ---
rests upon the unstated but implicit assumption that both quantities are of
roughly equal magnitudes.  Otherwise, this index reduces to a mere
measure of one quantity or the other, losing its hallmark sensitivity
to both quantities simultaneously.

In this paper, we have shown that this unfortunate outcome is the result of   
a missing step in the definition of $h$, namely the failure to introduce
a ``speed of light'' conversion factor within this comparison.
As we have seen, this feature, which is {\it required}\/ on dimensional grounds, 
automatically restores the $h$-index to its intended purpose,
rendering it invariant against arbitrary rescalings of either quantity
(citation counts or paper counts) with respect to the other.
Moreover, we have provided an explicit recipe whereby this missing
``speed of light''
can be calculated in order to guarantee that the resulting $h$-indices
are equally balanced against separate variations 
in the numbers of papers and the numbers of citations to those papers.

As indicated in the title and throughout the text of this paper,
we regard our introduction of the ``speed of light'' conversion factor
as a matter of {\it completing}\/, rather than modifying, the definition of the
$h$-index.  
We are not disagreeing with either the philosophy or the underlying
methodology according to which the $h$-index is constructed ---  we are merely 
supplying an important missing ingredient in 
order to ensure that the resulting index has all the mathematical
properties it logically should have. 
As such, we regard our proposal as compelled by internal logical necessity 
rather than a desire to somehow ``improve'' the $h$-index or extend its utility.
For this reason, we have not attempted to assess whether this 
modification to the $h$-index might actually prove useful for different communities
of scientists, choosing to defer this practical question for later investigation.
Rather, we regard this proposal as purely ``theoretical'' in the sense that
it advances a logical argument about
the internal self-consistency of the $h$-index itself and 
proposes a method
by which this mathematical self-consistency can be maintained.
To the best of our knowledge, such observations 
are new and do not appear anywhere in the prior bibliometrics literature.
This includes a rather large and impressive 
body of work~\cite{woeginger1,woeginger2,marchant1,marchant2,marchant3,
quesada1,quesada2,quesada3,quesada4,hwang,miroiu}
focusing on attempts to place the $h$-index on a solid axiomatic
footing as the inevitable bibliometric index meeting 
certain internal logical self-consistency criteria.

That said, there does exist a large prior literature addressing the purely
practical issue of {\it improving}\/ the $h$-index against a number of
perceived shortcomings.  A few representative papers 
are listed in Refs.~\cite{vanEck,waltman,Schreiberself,Egghe2,Komulski,Jin,Rons,
Schreiberhigh,Wu,Zhang,Petersen2,Petersen3,Batista,Egghe,Schreiber,Leydesdorff,hbar,Piazza,
Tawfik,Silagadze,Miskiewicz,Petersen,Iglesias,Zyl,Lundberg,Radicchi,RadicchiCast,RadicchiCast2,review}.
Proposals for modifying the $h$-index include
\begin{itemize}
\item  eliminating self-citations (or citations from collaborators or from scientific progeny) 
          from consideration~\cite{Schreiberself};
\item  increasing the weighting of very highly-cited papers, either through 
           the introduction of intrinsic weighting factors or the development of 
            entirely new indices
           which mix the $h$-index with other more traditional indices (such as total citation 
            count)~\cite{vanEck,waltman,Egghe2,Komulski,Jin,Rons,Schreiberhigh,Wu,Zhang,Petersen2,Petersen3};
\item considering the use of so-called ``fractional citations'' in which one divides
           the number of citations associated with a given paper by the number
              of authors on that paper~\cite{Batista,Egghe,Schreiber,Leydesdorff,hbar,Piazza};
\item considering the use of so-called ``normalized citations'' in which one
            divides each citation by the total number of papers in the reference
           list of the citing paper;
\item introducing correction factors which specifically compensate for the citation-related
         advantages accrued by authors who are members of large ``big science'' 
          collaborations~\cite{Tawfik};
\item  increasing the sensitivity to the variability (``entropy'') 
          of the overall citation profile of a given researcher~\cite{Silagadze};
\item   weighting the citations to a given paper according to the impact factor 
           of the journal in which the paper is published;
\item   weighting the citations to a given paper according to the impact factors 
           of the journals in which the {\it citing}\/ papers are published;
\item  differentiating between publications in peer-reviewed scientific journals 
           versus other forms of publications (proceedings articles, 
        book chapters, encyclopedia articles, {\it etc.}\/)~\cite{Miskiewicz}; 
\item   weighting the citations to a given paper according to its age, in order
          to eliminate the effects of time-dependent changes in citation patterns
           and community sizes~\cite{Petersen};
\item  weighting the citations to a given paper according to the relative age
          and/or status of the author in question compared with those of possible co-authors 
         on that paper (thereby recognizing the contributions of senior collaborators 
          in a different way than those of junior collaborators)~\cite{hbar};
\item  renormalizing the overall $h$-indices 
            within a given discipline in a discipline-dependent way~\cite{Iglesias,Zyl};
\item  renormalizing the number of papers of each scientist in a given discipline according to some average
          number of papers published by scientists in that discipline;  and
\item  renormalizing citation counts to each paper within a given discipline by the average number
            of citations per paper in that discipline~\cite{Lundberg,Radicchi,RadicchiCast,RadicchiCast2}.
\end{itemize}
Many other proposals exist as well (for a review, see Ref.~\cite{review}).
Needless to say, all of these proposals are made with one noble purpose in mind:
to produce a more ``just'' outcome for $h$ according to some particular sociological 
or ethical
measure of fairness, especially when comparing scientists in different disciplines.
Indeed, in some scientific disciplines the $h$-index can even fail 
to represent a fundamentally new bibliometric variable altogether, and is instead 
tightly correlated with a more traditional variable such as total citation count~\cite{redner,spruit}.

Such efforts at improving $h$ are certainly laudatory, as  indices such as $h$ often tend to play an inflated 
role in such practical matters as hiring decisions or determinations of grant sizes.
This is entirely understandable:  we live in a competitive world, 
and it is natural for players in that world to 
seek an apparently objective means of making decisions rather than
to rely solely on subjective impressions.
For these reasons, all of the above ideas can (and perhaps {\it should}\/) be 
pursued in relation to our ``completed'' $h$-index as well.
It may well turn out that
our ``completed'' $h$-index is even more amenable to 
the above sorts of further modifications than the traditional $h$-index, producing superior results.
However, 
no matter what the outcome of such future studies might be, 
we regard our ``completion'' of $h$ as fundamentally different from the 
above proposals in that
it is motivated not by issues of sociology, but rather by a need for internal mathematical 
consistency.
Even if situations are found in which
our ``completion'' of $h$ leads to results which
are inferior from a practical point of view, it would then still remain
for us to address the deeper theoretical question as to why 
those situations implicitly tend to prefer 
the rather arbitrary numerical value $c_A=1$.

Despite the difference in underlying motivations,
there are nevertheless certain operational similarities 
between our proposal and some of those listed above.
Perhaps the greatest similarity is with the final proposal in 
the list above, 
involving rescalings of citation counts.
In Refs.~\cite{Lundberg,Radicchi,RadicchiCast,RadicchiCast2},
it is argued that rescaling
each citation count $N_c$ by the average
number of citations per paper $N_c^{\rm (avg)}$ in a given discipline
allows a comparison between scientists in different disciplines.
If we take this proposal literally as a recipe for calculating 
a modified $h$ in each discipline,
the traditional $h$-condition becomes
\beq
 {N_c(h) \over N_c^{\rm (avg)}}~=~ h~.
\label{avg}
\eeq
At an algebraic level,
this equation is certainly similar to the $h$-defining condition (\ref{cond1}) we are proposing,
with a ``speed of light'' $c_A =  N_c^{\rm (avg)}$ for that discipline.

However, there are several critical differences between these proposals.
First and foremost, at a conceptual level, our $c_A$ is generally {\it not}\/ equal to the average number
of citations per paper in a given discipline --- $c_A$ is instead
a scale factor aimed at ensuring that the resulting $h$-indices are balanced,
exhibiting with equal sensitivities to paper counts and citation counts as far as possible.
Even worse, at a practical level, 
taking $c_A= N_c^{\rm (avg)}$ would result in individual $h$-values which 
are not only extremely unbalanced but also extremely small.
For example, for all of the profiles in Table~\ref{table1}, the $h$-index defined
according to Eq.~(\ref{avg}) would never exceed 4.
Since $h$ is restricted to be an integer, such an $h$-index would not rank scientists
so much as distribute them amongst only a few several rather large bins.

Finally, even if we were to weaken the condition in Eq.~(\ref{avg}) somewhat and 
consider dividing citation counts by some quantity $w^{(A)}$ which is merely {\it proportional}\/ to
$N_c^{\rm (avg)}$ with a discipline-independent proportionality constant,
this would still be functionally different from our proposal.  This follows from the fact, 
evident from the data in Table~\ref{table1}, that there is no direct relationship (proportional or otherwise)  
between $N_c^{\rm (avg)}$ and $c_A$ for different citation profiles.
Indeed, only if two communities happen to have citation profiles
which are multiples of each other will we find that $N_c^{\rm (avg)}\propto c_A$.
However, even in such cases, it is only for the specific choice $c_A$ --- rather than a rescaled
version $N_c^{\rm (avg)}$  --- that the resulting $h$-values will be properly balanced
between paper counts and citation counts.
Thus, to the best of our knowledge, our ``completion'' of the $h$-index is 
unique relative to the prior literature.
 
In general, there are many factors which
could potentially affect the value of $c_A$ within a given community.
Although community size is one obvious factor that we discussed throughout
this paper, there are also other factors such as the intrinsic
culture of the community as it pertains to the act of writing papers and citing
other papers.  For example, some communities may have a culture in which 
a scientist writes many short papers rather than a few long ones,
or in which a scientist cites relatively few other papers rather than
every other paper which ever commented on the subject at hand.
Other factors might include the inherent ``focus''  
of a community:  one community may have
members who tend to focus on a relatively small set of big 
questions, while another
community
may have members whose attention is more diffusely
distributed across an extremely diverse array of difficult problems.
Clearly, a community which is highly focused on  
fewer problems will tend to generate large numbers of 
cross-citations within its publications, even if its overall size is smaller.
However, as we have defined it,
$c_A$ captures all of these effects within
a single conversion factor.

Just as $c_A$ can depend on $A$,
it is important to realize that $c_A$ may also depend on {\it time}\/.
Scientific communities rarely remain static:  they evolve not only in size but
also in their intrinsic paper-writing and paper-citing cultures.
Thus, when we refer to a particular community as having a particular $c_A$,
we are implicitly referring to this community as it exists at a particular 
moment in time.
From this perspective, two communities which differ in scientific discipline but
exist at the same time
are no different from two communities which share the same scientific discipline
but exist in different eras --- 
indeed, in both cases each community may be treated 
as independent, distinct, and endowed with its own value of $c_A$.

In this paper, we have concentrated on the issue of 
``completing'' the definition of $h$ 
in such a way that $h$-indices remain balanced, as far
as possible, for scientists within a {\it single}\/ community.
As a result, only a single ``speed of light'' $c_A$ ever entered our calculations.
However, given the observations above,
we now discuss how our ``completed'' $h$-indices may be used in order to
compare scientists {\it across}\/ different communities.

In general, two different communities of scientists will have two different
collective citation profiles, $N_c^{(A)}$ and $N_c^{(B)}$.
These collective profiles will typically differ not only in their overall magnitudes but
also in their intrinsic shapes.
One therefore wonders whether it might be possible to somehow rescale 
these profiles in various ways in order to bring them into some degree of
alignment, all while
still preserving the fundamental meaning of $h$.

Viewed from this perspective, the recipe we have provided in this paper
can be interpreted as 
having the effect of rescaling citation counts $N_c(r)\to N'_c(r) \equiv N_c(r)/c$ 
by a certain ``speed of light'' $c$  
for each community in such a way that
\begin{itemize}
\item the $h$-defining condition $N'_c(h)=h$ now becomes a $45^\circ$ line (slope = $+1$) in 
     the rescaled $(r,N'_c)$ plane;  and
\item the rescaled $N'_c(r)$ profile function at this point also makes an angle of $45^\circ$ relative to the vertical (slope = $-1$) in the rescaled $(r,N'_c)$ plane.
\end{itemize}
Indeed, the fact that both angles will be $45^\circ$
when plotted versus rescaled citation counts
is directly related to the ``equal angles'' requirement in Fig.~\ref{fig:spacetime2}.
Thus, expressed in terms of rescaled citation counts,
we see that our recipe for calculating the ``speed of light'' $c$ 
is precisely that which turns Figs.~\ref{fig:spacetime}(b) and
\ref{fig:spacetime}(c) into Fig.~\ref{fig:spacetime}(a), as far as possible, ensuring a balanced value of $h$.
Thus, if we perform this rescaling
separately for $N_c^{(A)}$ and $N_c^{(B)}$,
each with its own ``speed of light'' $c_A$ and $c_B$ respectively,
we are then guaranteed that our rescaled citation profiles
$N_c^{\prime (A)}(r)$ and $N_c^{\prime (B)}(r)$ will be ``aligned'' insofar as
they will now both share the two characteristics listed above. 

However, even after these rescalings are performed,
there remains one additional possible rescaling 
between $N_c^{\prime (A)}(r)$ and $N_c^{\prime (B)}(r)$ 
for which we have not yet accounted:  unlike rescalings of citation counts {\it relative}\/ to paper counts,
there is also the possibility of rescaling {\it both}\/ of these variables by a common factor $a$.
At an algebraic level, this final possible rescaling corresponds to an overall 
origin-centered dilation 
within the $(r,N'_c)$ plane --- \ie, a magnifying transformation 
such as $N_c(r)\to  a N_c(r/a)$ 
which leaves $c_A$
invariant but which nevertheless rescales $h_A$ by the factor $a$.
For this, however, the solution is simple:  we simply
rescale the $h$-indices within each community 
in such a way that $h_A$ and $h_B$ are made equal.
Indeed, this is where the quantities $h_A$ and $h_B$ (which emerged
as by-products of our calculations of the ``speed of light'' factors $c_A$ and $c_B$)
prove useful.
Geometrically, this final rescaling of the
citation profiles for Communities~$A$ and $B$ 
has the effect of ensuring that these profile functions
not only share a common {\it slope}\/ along the $45^\circ$ line,
but now also share a common {\it point}\/ along the $45^\circ$ line.
Thus, in this way, we have ensured that our two community citation profiles are as closely aligned as possible
in the region of greatest significance for $h$.
Of course, if Community~$B$ is simply an ensemble of multiple copies of Community~$A$,
then we will find $h_A=h_B$ automatically.
In such cases no further rescaling will be needed.

Thus, combining these two sets of rescalings, we see that our general procedure for comparing
scientists across different communities $A$ and $B$ is relatively simple.  
Let us assume that
Scientist~$X$ is a member of Community~$A$
and that Scientist~$Y$ is a member of Community~$B$.
Let us further assume that 
Communities~$A$ and $B$ have citation profiles $N_c^{(A,B)}(r)$
respectively, and
that Scientists~$X$ and $Y$ have profiles $N_c^{(X,Y)}(r)$ 
respectively.
Following the procedure outlined above,
we determine $h^{(X)}$ by setting
$N_c^{(X)}(h^{(X)})=c_A h^{(X)}$,
and $h^{(Y)}$ by setting
$N_c^{(Y)}(h^{(Y)})=c_B h^{(Y)}$.
In these relations, the ``speeds of light'' $c_A$ and $c_B$ are calculated as 
solutions
to the simultaneous constraint equations~(\ref{con1}) and (\ref{con2}),
using the community profiles $N_c^{(A)}$ and $N_c^{(B)}$ respectively in these
equations.
We then rescale $h^{(X)}$ and $h^{(Y)}$ according to the
respective values $h_A$ and $h_B$ which also emerge from
these calculations.
If $h^{(X)}/h_A > h^{(Y)}/h_B$, we then conclude 
that the relative $h$-based publication impact of Scientist~$X$ exceeds
that of Scientist~$Y$.
Indeed, while there may well exist other ways of reaching
a similar conclusion using other impact measures,  
we believe that the above
procedure is the only mathematically 
consistent way of using $h$-indices to conduct such interdisciplinary comparisons 
while simultaneously remaining true to the underlying balance-based approach to $h$. 

In this connection, we note that the approach advocated in Ref.~\cite{Iglesias}
for such inter-community comparisons
(an approach which corresponds to rescaling all of the $h$-indices
of the researchers in a given community by a common community-dependent average number of citations 
per paper)
is reminiscent of the final rescaling that we performed above --- \ie,  
our origin-centered dilation.  However, the approach we have outlined above
has three critical differences
relative to that of Ref.~\cite{Iglesias}.
First, the corresponding rescaling  
factor is not given by an average number of citations per paper,
but rather by the quantity $h_A/h_B$, to which it will in general bear no relation.
Second, our prior rescalings 
were also critical to our analysis, ensuring that the rescaled aggregate community 
profiles $N_c^{(A,B)}(r)$ will continue to satisfy the two crucial conditions  
bulleted above.
Finally, the $h$-indices for each member of each community must be calculated
as we have advocated all along, with reference to an appropriate ``speed of light''. 
Indeed, it is only through such a step-by-step procedure that the mathematical self-consistency 
of such an inter-community comparison can be maintained.

Finally, we remark that there may exist scientists who are simultaneously members of 
multiple, otherwise disjoint research communities.  
For example, a high-energy physicist might occasionally write an 
article on bibliometrics.   Such a  scientist will then have 
multiple disjoint sets of publications, and to each set we 
may associate its own $h$-index calculated with its own ``speed 
of light'' $c_A$.   In other words, for all intents and purposes, 
such a scientist functions no differently than do multiple scientists 
who share a common identity but otherwise lead parallel independent careers 
within disjoint communities.  At first glance, it may seem that this 
dichotomy misses something essential in that it fails to reward the breadth 
of the researcher in question while nevertheless potentially penalizing
the researcher for his/her reduced (partitioned) productivity within 
each community.   However, the sad fact is that one cannot enhance 
one's impact in high-energy physics by writing articles on bibliometrics, 
no matter how brilliant those bibliometrics articles might be --- 
only articles on high-energy physics can do that.   
Such individuals can nevertheless take solace in having produced a body of work 
with multiple, independent $h$-factors along multiple lines of research.
Indeed, the research output of such individuals may be characterized as 
having $h$-indices which are {\it vectors}\/ rather than scalars.
It is only due to the limitations of time and space that we refrain from contemplating  
the numerous geometric ramifications of this observation at this juncture.


\begin{acknowledgments}


This paper was conceived and written in the wee hours that ordinarily should be
spent on sleep.
As such, it  bears absolutely no relation to the research or other professional activities
of the author,
and moreover does not represent the opinions or conclusions of any funding agency  
whatsoever.
That said, the normal research activities of the author are
funded in part under DOE Grant DE-FG02-13ER-41976.~
Useful comments and encouragement from
S.~Redner,
M.~Sher, S.~Su, B.~Thomas, L.~Waltman, and J.~Wells 
are also herewith gratefully acknowledged.

\end{acknowledgments}


\end{document}

\end{references}

\end{document}